# Recent Reversal in Single-Authored Papers

Abdulkareem Alsudais

College of Computer Engineering and Sciences, Prince Sattam bin Abdulaziz University, Al-Kharj 11942, Saudi Arabia
E-mail: A.Alsudais@psau.edu.sa

## Abstract

The number of authors per scientific paper has risen for decades, and the share of single-authored papers has correspondingly declined, a pattern long documented across fields. This study shows that on arXiv.org this decline has recently reversed. Findings indicate that for submissions between 2015 and early 2026, after falling steadily for a decade, the single-author share rose from 10.9% in 2024 to 11.2% in 2025, and in the first quarter it rose from 10.7% in 2025 to 12.9% in 2026. This is the largest single year-over-year first-quarter change in the studied window, and the first increase in the first-quarter series. In 2025, for the first time in the studied period, the number of single-authored papers grew faster year-over-year than the number of multi-authored papers, reversing the long-standing pattern. This reversal is notable because it is broad, rising within all eight top-level arXiv categories, and because it reflects a genuine change in authoring practice within fields rather than the shifting disciplinary composition of arXiv submissions. A comparison of five arXiv snapshots spanning 2022 to 2026 confirms that the finding is not an artifact of incomplete or revised author metadata in recent submissions. As robustness checks, article-type records from OpenAlex, a database covering a broader range of disciplines, and an arXiv snapshot extended through September 2026 both confirm these findings. This latest snapshot shows the share still rising, to 14.1% in Q2 and 17.0% in Q3 of 2026, compared with 10.5% and 11.1% in the same quarters of 2025.

***Keywords**:* Authorship trends; Scientific writing; Research collaboration;

# 1. Introduction

The number of authors per paper has been a subject of study for decades, with research examining how team sizes vary across fields, how they relate to citation impact, and how they have changed over time (Bornmann & Osório, 2019; Jakab et al., 2024; Lin et al., 2025). A substantial body of work has documented this long-term shift toward team-based research and the corresponding decline in single-authored papers across disciplines (Erfanmanesh & Torabi, 2026; Larivière et al., 2015; Praus, 2025). At the same time, the recent rise and popularity of AI models and tools that can assist in scientific research have prompted research into the impact of AI on many aspects of scientific publishing (Alsudais, 2025a, 2025b; Kousha & Thelwall, 2026; Lehr et al., 2024).

This study is motivated by the possibility that the rapid adoption of accessible generative AI tools has reshaped how scholars produce papers, and in particular that it may be associated with a rise in the share of single-authored work. Recent studies have highlighted the benefits of AI usage on the productivity and efficiency of researchers (Dell'Acqua et al., 2026; Kusumegi et al., 2025; Madanchian & Taherdoost, 2025; Mohammadi et al., 2026). Additionally, a recent preprint introduces the notion of "vibe researching," defining it as "a mode of scientific inquiry in which a human researcher provides high-level direction, creative intuition, and critical evaluation, while LLM-based agents carry out the labor-intensive parts of the process (literature discovery, experimental implementation, data analysis, manuscript preparation) in response to natural language instruction" (Feng & Liu, 2026). To the extent that such tools reduce a researcher's dependence on co-authors, whether for drafting, language assistance, analysis, or other tasks once shared among collaborators, an increase in single-authored papers would be a plausible consequence.

The aim of this study is to analyze whether such a change is observable in the data and how it varies across scientific fields, without attempting to identify the specific causes of any observed patterns, including any role of AI. Because the dataset used in this study comprises eight primary categories whose relative shares have changed over time, the study also examines whether any observed change arises from shifts within categories or from changes in the overall composition of submissions to the platform.

Finally, the study draws on five snapshots of the dataset to assess whether any observed change could instead reflect subsequent revisions to author counts, for example papers initially posted by a single author that later gained additional authors.

In this paper, the objective is to address the following research questions:

- RQ1: How has the share of single-authored papers on arXiv evolved between 2015 and early 2026, and is there evidence of a departure from the long-run trend in the most recent period?
- RQ2: To what extent are observed changes in the aggregate single-author share associated with changes in single authorship within fields, as opposed to changes in the composition of arXiv submissions across fields?
- RQ3: Do the observed trends, and any recent departure from them, differ across the eight top-level categories on arXiv?

## 2. Dataset

| Snapshot ID | Date Posted | Number of Papers |
|---|---|---|
| 72 | April 10, 2022 | 2,047,905 |
| 124 | April 16, 2023 | 2,238,880 |
| 175 | April 14, 2024 | 2,459,562 |
| 227 | April 13, 2025 | 2,710,806 |
| 281 | April 12, 2026 | 3,008,380 |

**Table 1.** The five snapshots used in the study, with their upload dates, and number of papers.

To investigate the research questions, arXiv.org is used as the main source of papers. According to its official description, “arXiv is a curated research-sharing platform open to anyone. As a pioneer in digital open access, arXiv.org now hosts more than three million scholarly articles in eight subject areas” (arXiv, 2026). The eight categories papers can belong to are: physics, mathematics (Math), computer science (CS), quantitative biology (q-bio), quantitative finance (q-fin), statistics (Stat), electrical engineering and systems science (EESS), and economics (Econ). Since 2020, the entire metadata corpus has been available for download through the dataset and competition hosting platform Kaggle. The corpus is updated approximately weekly and is distributed as a JSON file.

To assess the stability of metadata over time and to validate the analyses presented in subsequent sections, five snapshots of the Kaggle distribution are used. These are version 72, version 124, version 175, version 227, and version 281. Each snapshot was downloaded from Kaggle directly. The five snapshots span approximately four years and are roughly equally spaced. The primary analyses in this paper use version 281, the most recent of the five snapshots, which contains all submissions through early April 2026. Table 1 provides details on the five snapshots. A later snapshot, version 306 (posted on October 3, 2026, and containing 3,195,172 papers), is used only in the robustness check in Section 10. That check extends the analysis through September 2026 to validate the main results.

The primary reason for using these five snapshots is to examine whether the observed increase in single-authored papers in 2025 and early 2026 might be an artifact of authors adding co-authors to their papers after the initial posting. Each snapshot contains all papers posted on the platform as of the snapshot date and earlier snapshots remain available for download. Papers on arXiv can also have multiple versions, allowing authors to update information about a paper, such as its abstract, title, or content, after the initial posting. Each paper belongs to one or more primary categories. For example, a paper may be posted under both the computer science and the mathematics categories.

Every paper has at least one version. For this study, the upload date of the first version (v1) is used to determine the submission year, and where required, the submission month. This choice differs from the method used in some prior work that uses the most recent version date, and reflects the central purpose of this study: analyzing the moment at which authoring decisions, including the decision to publish alone or with co-authors, were first made. Subsequent revisions can add or remove authors, but these revisions are themselves the subject of the analysis of author-list stability described in Section 8. The other preprocessing steps used here to identify categories and versions follow those used in prior work, which can be accessed for further detail (Alsudais, 2025a).

## 3. Author Count Determination

An important processing step for this study is to extract the number of authors for each paper. The dataset provides both a parsed author list (“authors_parsed”) and an unparsed author string (“authors”). The parsed list is used to determine the number of authors, by counting the number of entries it contains. To confirm the accuracy of this approach, a random sample of 385 papers was drawn (95% confidence, ±5% margin of error) and each was manually checked. In every case, the author count derived from the parsed list matched the number of authors in the paper. In two cases, however, one of the listed “authors” was not an individual but the name of an institution or research group.

Further analysis was conducted to assess whether such instances are common. This analysis involved creating a simple rule-based classifier to identify when a listed author is, or writes on behalf of, an institution or research group, validating the classifier’s performance, and rerunning the entire analysis using a stricter definition of “single author,” under which such papers no longer count as single-authored. The classifier ignores capitalization and applies two rules. It flags an author entry if the name contains the word “collaboration” or “consortium” (e.g., “ATLAS Collaboration”), or if “for” or “on behalf of” is followed by “collaboration,” “group,” “team,” or “consortium,” indicating an individual writing on behalf of a group (e.g., “Burt Holzman (for the PHOBOS Collaboration)”). To verify the classifier, a sample of 250 single-authored papers, stratified by the classifier’s label, was annotated manually without knowledge of the classifier’s decisions. The sample deliberately over-represented papers the classifier labeled as collaborations, so that errors in both directions could be detected. The manual annotation agreed with the classifier’s label in every case, which supports applying the classifier to the entire dataset.

The proportion of single-authored papers that in fact represent a collaboration was small in every year, between 1.3% and 2.7%, and was lower in 2025 (1.5%) and January to early April 2026 (1.3%) than in 2015 (2.7%) and 2016 (2.2%). Because the stricter definition removes a larger proportion of single-authored papers in 2024 (1.8%) than in 2025 (1.5%) or January to early April of 2026 (1.3%), it marginally strengthens rather than weakens the reversal. The increase in the single-authored share from 2024 to 2026 is 2.11 percentage points under the strict definition, compared with 2.08 under the original definition. Given that the effect on the results is negligible, and to keep the definition of a single-authored paper simple and easily reproducible, the analyses reported in this paper classify a paper as single-authored whenever its author list contains exactly one entry, regardless of whether that entry is an individual or a collaboration. Appendix A provides additional details of the classifier, including examples (Table A.1), performance metrics, and the year-by-year effect of reclassification (Table A.2).

## 4. Papers Count and Category Composition

Two ways of counting papers are used in this study, depending on the level of analysis. In the aggregate analysis, each paper is counted exactly once, regardless of how many categories it belongs to. The aggregate single-author share for a given year is then the number of distinct single-authored papers divided by the number of distinct papers with at least one author.

In the category-level analysis, a paper assigned to more than one top-level category is counted once within each of those categories. For example, a paper assigned to both computer science and mathematics contributes to the single-author rate of computer science and to the single-author rate of mathematics. This is the appropriate treatment for category-level rates, since such a paper genuinely belongs to both fields, and assigning it to only one would distort the rates of both. A consequence of this choice is that the aggregate single-author share obtained by summing across categories differs slightly from the share obtained by counting each paper once.

A further consideration arises when interpreting changes in the aggregate single-author share over time. The composition of arXiv submissions has changed substantially over the study period, with some categories (notably computer science) growing far faster than others. Because categories differ in their typical rates of single authorship, a change in the aggregate share can arise from two sources: a genuine change in authoring practice within fields, or a change in how papers posted to arXiv are distributed across categories. If a category with a low single-author rate grows as a share of all submissions, the aggregate share would fall even if no category changed its own rate. Distinguishing the two sources is important, since an aggregate trend driven by compositional change carries a very different interpretation from one driven by within-field behavior.

To address this, the analysis reports the single-author share at two levels. At the aggregate level, the overall single-author share is tracked across all submissions for each year. At the category level, the within-category single-author share is tracked separately for each of the eight top-level categories. Examining the within-category shares directly makes it possible to determine whether a change in the aggregate share is accompanied by changes within the individual categories. If the single-author share declines or rises within most categories considered separately, the aggregate change cannot be attributed solely to the changing composition of arXiv submissions, since the change is also present within fields whose composition is held fixed when the category is examined on its own.

To quantify the contribution of compositional change directly, a simple counterfactual is computed. For a given pair of years, the aggregate single-author share that would have obtained in the later year is calculated under the assumption that the distribution of papers across the eight categories had remained fixed at its earlier-year level while each category's own single-author rate took its observed later-year value. The counterfactual share is therefore obtained by weighting each category's later-year rate by its earlier-year share of submissions and summing across categories. The difference between this counterfactual share and the observed earlier-year share reflects within-category rate change in the absence of compositional change, while the difference between the counterfactual share and the observed later-year share reflects the additional contribution of compositional change. The two contributions sum exactly to the observed change. This calculation uses only the category-level shares and rates already reported and introduces no assumptions beyond the fixed-composition scenario. Because category-level rates are required, the counterfactual uses the category-level counting described above.

## 5. The Trend in Single-Authored Papers (RQ1)

| Year | Total | Single | Multi | Single share (%) | Multi share (%) | Single growth (%) | Multi growth (%) | Total growth (%) | Single Share change (pp) |
|---|---|---|---|---|---|---|---|---|---|
| 2015 | 105,130 | 21,637 | 83,493 | 20.58 | 79.42 | – | – | – | – |
| 2016 | 113,440 | 21,830 | 91,610 | 19.24 | 80.76 | +0.89 | +9.72 | +7.90 | −1.34 |
| 2017 | 123,781 | 22,293 | 101,488 | 18.01 | 81.99 | +2.12 | +10.78 | +9.12 | −1.23 |
| 2018 | 140,377 | 22,811 | 117,566 | 16.25 | 83.75 | +2.32 | +15.84 | +13.41 | −1.76 |
| 2019 | 155,917 | 23,180 | 132,737 | 14.87 | 85.13 | +1.62 | +12.90 | +11.07 | −1.38 |
| 2020 | 178,275 | 24,763 | 153,512 | 13.89 | 86.11 | +6.83 | +15.65 | +14.34 | −0.98 |
| 2021 | 181,599 | 23,895 | 157,704 | 13.16 | 86.84 | −3.51 | +2.73 | +1.86 | −0.73 |
| 2022 | 185,987 | 23,566 | 162,421 | 12.67 | 87.33 | −1.38 | +2.99 | +2.42 | −0.49 |
| 2023 | 209,223 | 25,149 | 184,074 | 12.02 | 87.98 | +6.72 | +13.33 | +12.49 | −0.65 |
| 2024 | 243,690 | 26,581 | 217,109 | 10.91 | 89.09 | +5.69 | +17.95 | +16.47 | −1.11 |
| 2025 | 284,161 | 31,850 | 252,311 | 11.21 | 88.79 | **+19.82** | +16.21 | +16.61 | **+0.30** |
| *2026** | *85,475* | *11,101* | *74,374* | *12.99* | *87.01* | – | – | – | ***+1.78*** |

**Table 2.** Annual single- and multi-authored arXiv submissions, 2015–2026. *Partial data for 2026

| Year | Total | Single | Multi | Single share (%) | Multi share (%) | Single growth (%) | Multi growth (%) | Total growth (%) | Single share change (pp) |
|---|---|---|---|---|---|---|---|---|---|
| 2015 | 24,914 | 5,263 | 19,651 | 21.12 | 78.88 | – | – | – | – |
| 2016 | 27,211 | 5,379 | 21,832 | 19.77 | 80.23 | +2.20 | +11.10 | +9.22 | −1.36 |
| 2017 | 28,975 | 5,338 | 23,637 | 18.42 | 81.58 | −0.76 | +8.27 | +6.48 | −1.34 |
| 2018 | 32,605 | 5,557 | 27,048 | 17.04 | 82.96 | +4.10 | +14.43 | +12.53 | −1.38 |
| 2019 | 35,964 | 5,592 | 30,372 | 15.55 | 84.45 | +0.63 | +12.29 | +10.30 | −1.49 |
| 2020 | 39,350 | 5,808 | 33,542 | 14.76 | 85.24 | +3.86 | +10.44 | +9.41 | −0.79 |
| 2021 | 43,799 | 5,882 | 37,917 | 13.43 | 86.57 | +1.27 | +13.04 | +11.31 | −1.33 |
| 2022 | 44,776 | 5,815 | 38,961 | 12.99 | 87.01 | −1.14 | +2.75 | +2.23 | −0.44 |
| 2023 | 46,887 | 5,850 | 41,037 | 12.48 | 87.52 | +0.60 | +5.33 | +4.71 | −0.51 |
| 2024 | 57,949 | 6,401 | 51,548 | 11.05 | 88.95 | +9.42 | +25.61 | +23.59 | −1.43 |
| 2025 | 64,850 | 6,956 | 57,894 | 10.73 | 89.27 | +8.67 | +12.31 | +11.91 | −0.32 |
| 2026 | 77,490 | 10,020 | 67,470 | 12.93 | 87.07 | **+44.05** | +16.54 | +19.49 | **+2.20** |

**Table 3.** First quarter single- and multi-authored arXiv submissions, 2015–2026.

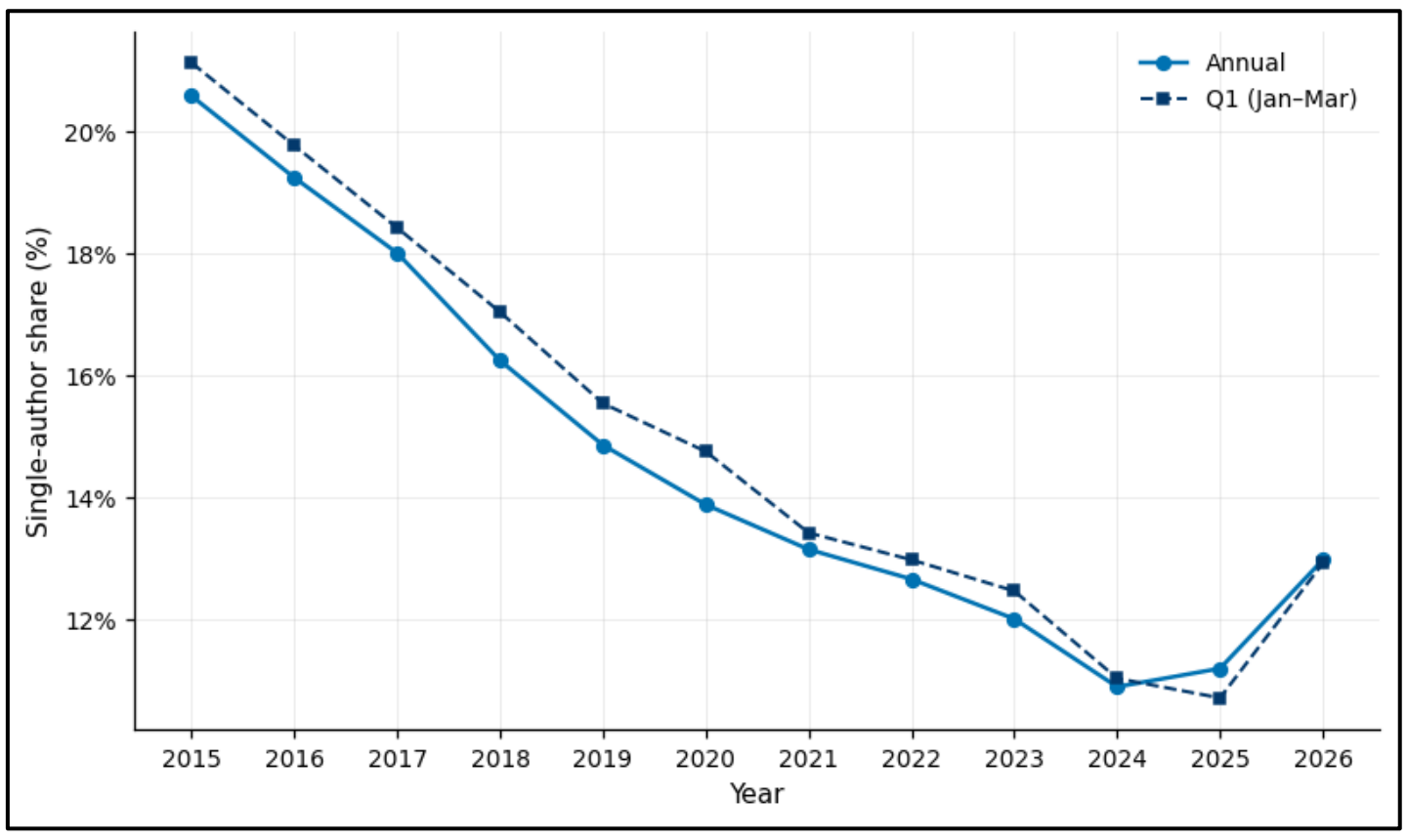

**Figure 1.** Single-author share of arXiv submissions, 2015 to 2026, shown for the full year and for the first quarter (January to March) of each year, with the 2026 full-year value reflecting only submissions posted through early April 2026.

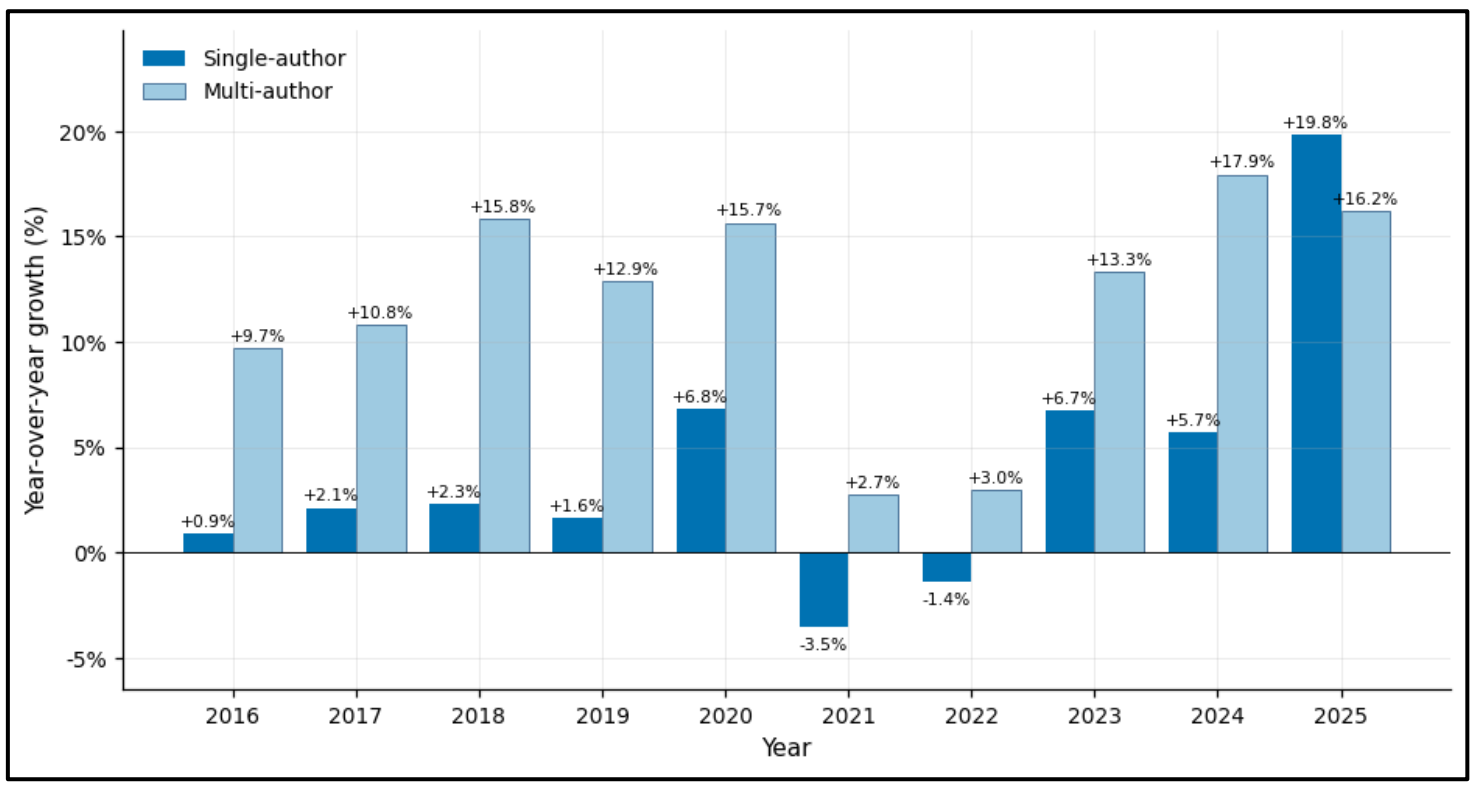

**Figure 2.** Year-over-year growth in the number of single- and multi-authored arXiv submissions, annual data, 2016 to 2025, with 2026 omitted because it is a partial year.

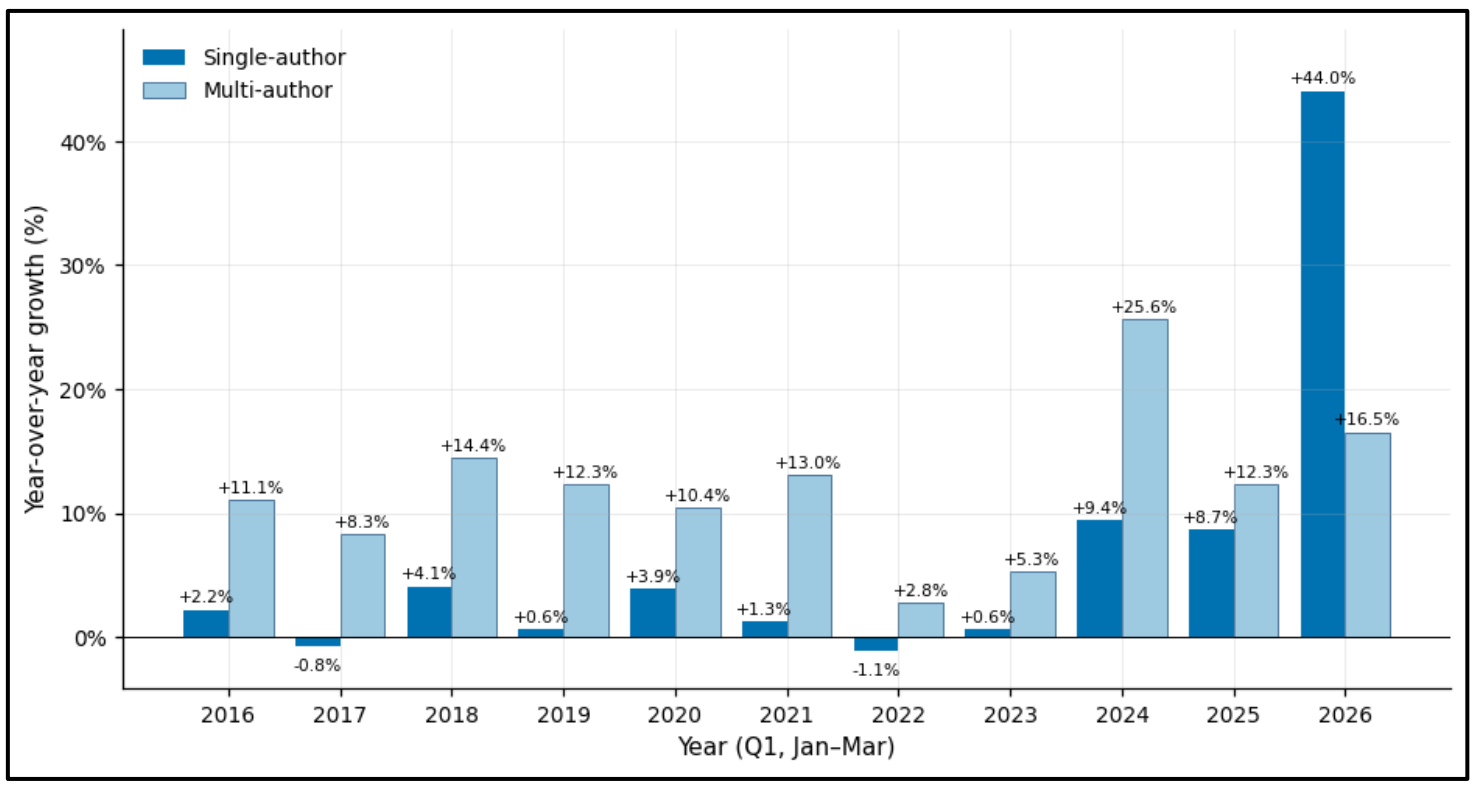

**Figure 3.** Year-over-year growth in the number of single- and multi-authored arXiv submissions, first-quarter (January to March) data, 2016 to 2026.

The first research question (RQ1) concerns how the share of single-authored papers on arXiv has evolved between 2015 and early 2026, and whether there is evidence of a departure from the trend in the most recent period. The year 2015 was chosen as the starting point to provide an eleven-year window. Because the snapshot used for the primary analyses (version 281) was taken in April 2026, the data for 2026 covers only approximately the first three and a half months of the year. To enable a like-for-like comparison unaffected by this partial coverage, a parallel analysis is performed at the first-quarter level: the single-author share among papers submitted in January through March is computed for each year from 2015 through 2026.

For each year, the aggregate single-author share is computed at the distinct paper level, and year-over-year changes are reported as percentage-point differences. In addition, the year-over-year growth in the number of single-authored papers is compared with that of multi-authored papers, since a shift toward single authorship would be reflected in single-authored papers growing more quickly. Because the 2026 data cover only a partial year, this growth comparison is reported at the annual level for complete years and at the first-quarter level, where 2026 is fully observed.

The results show that in 2015, 20.58% of all submissions were single-authored. By 2024 this share had fallen to 10.91%, before rising to 11.21% in 2025. Tables 2 and 3 present the annual and first-quarter figures, each reporting submission counts, single- and multi-author totals and shares, and year-over-year changes. Figure 1 presents the single-author share over time, for both the full year and the first quarter. The results show two distinct phases. From 2015 through 2024, the single-author share declined in every year, falling from 20.58% to 10.91% at the annual level and from 21.12% to 11.05% in the first quarter. Throughout this period the number of single-authored papers generally continued to rise, but the number of multi-authored papers rose faster, so that the single-author share fell year after year. This is reflected in the year-over-year growth columns of Tables 2 and 3.

This pattern reversed in the most recent period. At the annual level, the single-author share rose by 0.30 percentage points between 2024 and 2025, the first increase in the studied window, and the partial data for 2026 indicate a further rise to 12.99%. In the first quarter, the share rose from 10.73% in 2025 to 12.93% in 2026, an increase of 2.20 percentage points and the largest year-over-year first-quarter change, in either direction, observed across the period. This indicates a substantive shift rather than a marginal fluctuation.

The reversal is also evident in the relative growth of single- and multi-authored papers. Figure 2 presents the year-over-year growth of both at the annual level, and Figure 3 presents the same for the first quarter. In every year from 2016 through 2024, multi-authored papers grew more quickly than single-authored papers. In 2025, this pattern reversed for the first time at the annual level. Single-authored papers grew by 19.82%, exceeding the 16.21% growth of multi-authored papers. The same reversal is observed in the first-quarter data in 2026, and more strongly, with single-authored papers growing by 44.05% relative to the first quarter of 2025, compared with 16.54% growth in multi-authored papers. For the first time in the studied window, single-authored work was the faster-growing component of arXiv submissions. Two-proportion z-tests indicate that these changes exceed normal year-to-year variation: every annual decline from 2016 through 2024 is statistically significant ($p < 0.0001$), as are the 2025 increase (+0.30 percentage points; $p < 0.001$) and the first-quarter increase between 2025 and 2026 (+2.20 percentage points; $p < 0.001$). Appendix B describes the method and reports the full results.

The results reveal several findings. First, the share of single-authored papers declined steadily over the decade from 2015 to 2024, consistent with the long-documented shift toward team-based research. Second, this decline reversed in the most recent period, with the single-author share rising at the annual level in 2025 and in the first quarter in 2026, and with single-authored papers becoming the faster-growing component for the first time in the studied window. Third, while the magnitude of the reversal is modest relative to the cumulative decade-long decline, it is unprecedented within the studied window in its direction.

# 6. Within-Field Change versus Compositional Change (RQ2)

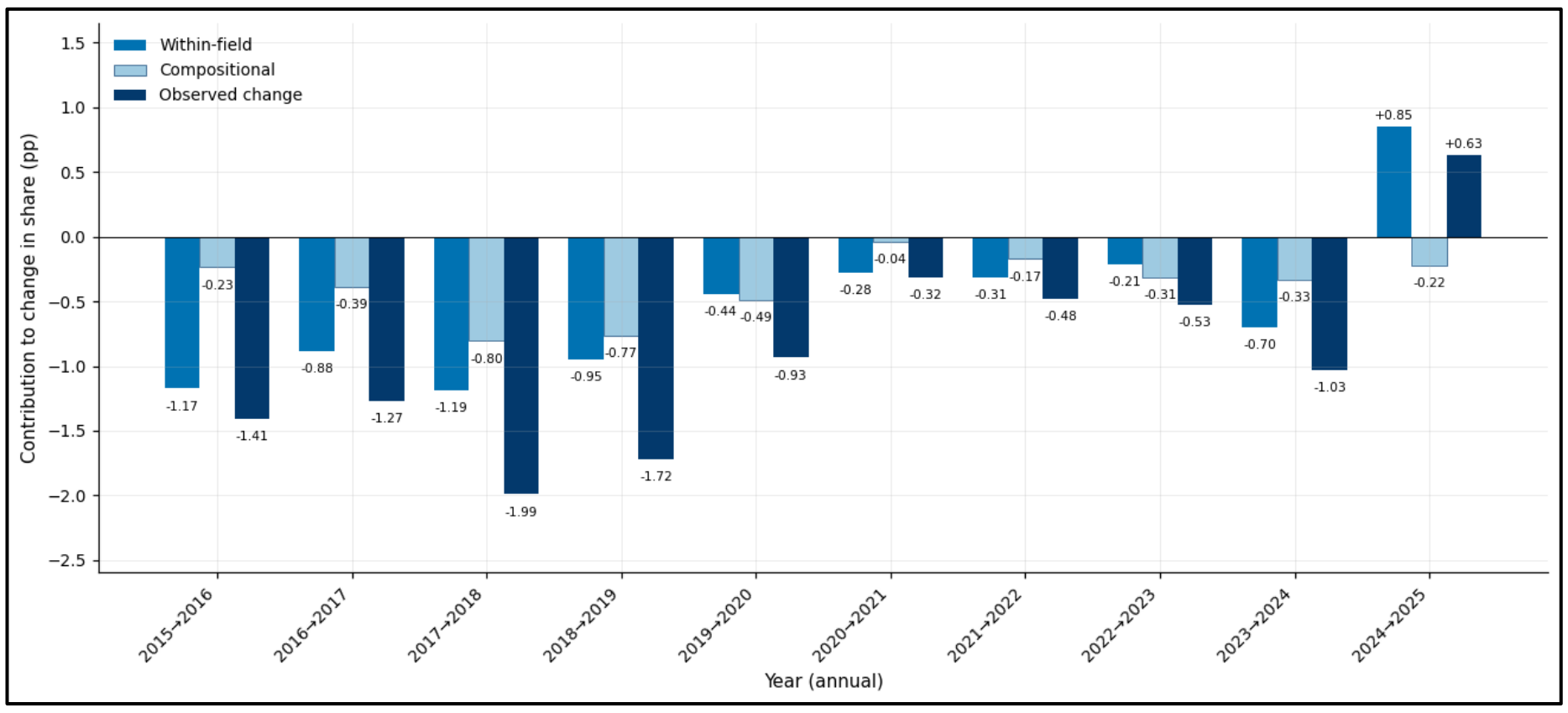

**Figure 4.** Year-over-year decomposition of the change in the aggregate single-author share, annual data, 2015 to 2025, in percentage points.

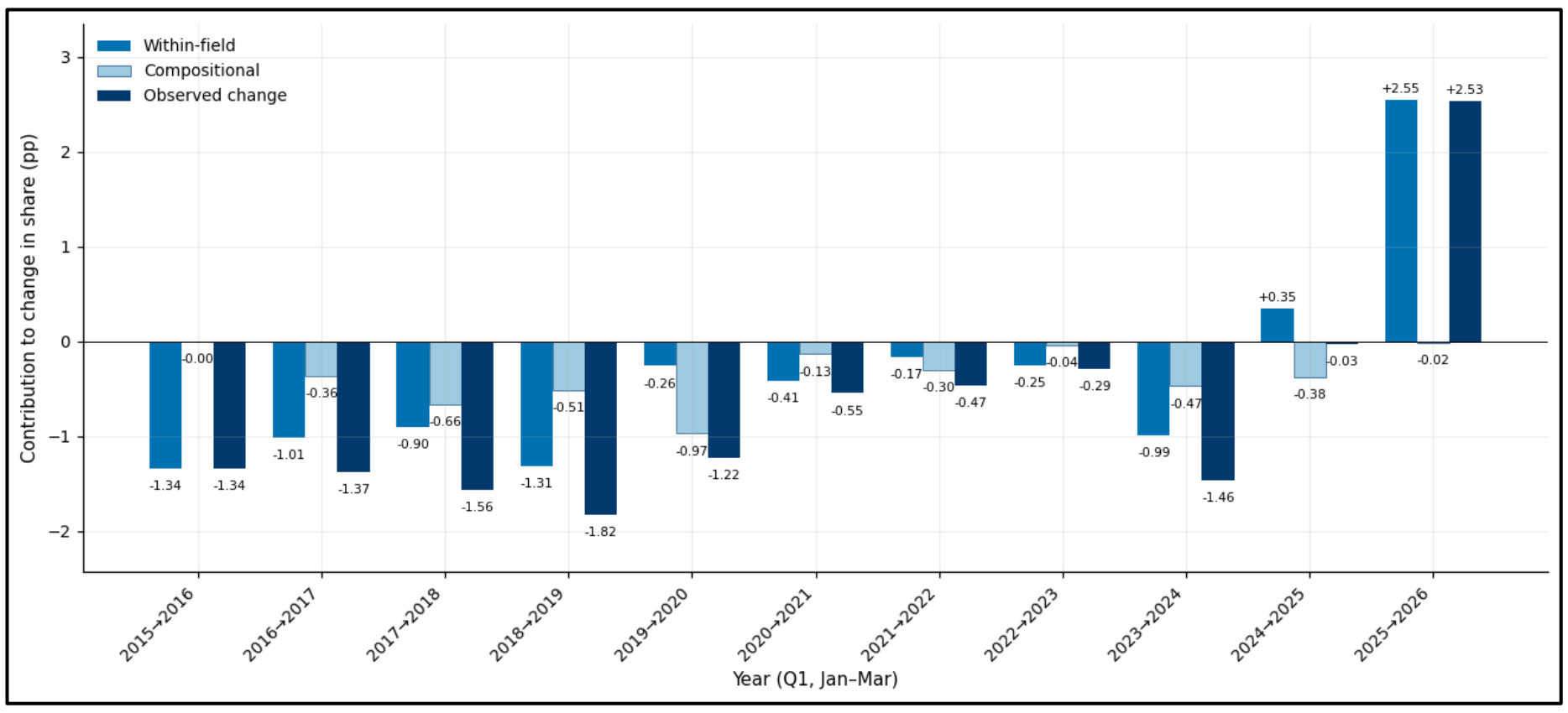

**Figure 5.** Year-over-year decomposition of the change in the aggregate single-author share, first-quarter (January to March) data, 2015 to 2026, in percentage points.

The second research question (RQ2) is about the extent to which changes in the aggregate single-author share reflect changes within fields rather than changes in the composition of arXiv submissions across fields. The within-category single-author shares are reported alongside the aggregate share, and the fixed-composition counterfactual described in Section 4 is computed for the full window and for the most recent year-over-year transitions. This makes it possible to state, for each comparison, what portion of the observed change is associated with within-category change and what portion with compositional change.

Over the full window, the decline was predominantly within-field. Between 2015 and 2025 the single-author share fell by 9.04 percentage points, from 20.40% to 11.36%. Within-field rate changes account for 5.57 percentage points of this decline (approximately 62%), and the shift in composition toward fields with lower single-author rates, most notably computer science, accounts for the remaining 3.47 percentage points. The first-quarter analysis gives a closely similar split of approximately 64% within-field. Alternative weighting conventions, such as fixing the reference composition at the later year or averaging the two endpoints, change the precise split only modestly and leave within-field change the larger contributor.

The recent reversal is even more clearly within-field. Between the first quarter of 2025 and the first quarter of 2026, the share rose by 2.53 percentage points, of which within-field change contributed 2.55 and compositional change contributed −0.02. The same pattern holds at the annual level, where the within-field component between 2024 and 2025 was +0.85 against a compositional component of −0.22.

Figures 4 and 5 present the year-over-year decomposition for the annual and first-quarter series. Across every year of the studied period the compositional component is negative, reflecting the steady growth of fields with lower single-author rates, while the within-field component is negative throughout the decline and turns positive in 2024–2025. In the first-quarter data it turns positive in 2025, a year before the first-quarter share itself rises.. The continued growth of computer science therefore acted weakly against the recent increase rather than driving it, and the reversal cannot be attributed to a change in the disciplinary composition of arXiv submissions.

## 7. Variation Across Categories (RQ3)

| Category | 2016 | 2017 | 2018 | 2019 | 2020 | 2021 | 2022 | 2023 | 2024 | 2025 | 2026* |
|---|---|---|---|---|---|---|---|---|---|---|---|
| CS | -2.45 | -1.74 | -1.46 | -0.98 | -0.48 | -0.42 | -0.60 | -0.06 | -0.66 | 0.88 | 2.04 |
| Physics | -1.01 | -0.34 | -1.11 | -0.55 | -0.84 | -0.08 | -0.10 | -0.27 | -0.75 | 0.09 | 1.93 |
| Math | -0.63 | -1.04 | -1.09 | -1.23 | -0.27 | -1.16 | -0.13 | -0.61 | -0.77 | 1.15 | 1.61 |
| Stat | -2.01 | -1.74 | -2.03 | -1.12 | 0.00 | 1.71 | 0.38 | -0.93 | 0.32 | 2.13 | 3.05 |
| EESS | 1.21 | 1.71 | -0.80 | -1.58 | -0.19 | 0.13 | -0.87 | 0.29 | -0.56 | 0.99 | 1.50 |
| q-bio | -1.54 | -0.82 | -0.52 | -1.90 | 1.37 | -1.90 | -0.54 | 0.15 | -0.89 | 0.59 | 5.75 |
| q-fin | -0.90 | -3.58 | 2.29 | -4.53 | 2.85 | -1.00 | -0.24 | 2.37 | -3.54 | 7.46 | 2.43 |
| Econ | -5.11 | 4.49 | 9.59 | -5.92 | 0.96 | 1.85 | -0.28 | 1.77 | -2.52 | 5.73 | 0.35 |

**Table 4.** Year-over-year change in the within-category single-author share by category, annual, 2016 to 2026, in percentage points. *Partial data for 2026.

| Category | 2016 | 2017 | 2018 | 2019 | 2020 | 2021 | 2022 | 2023 | 2024 | 2025 | 2026 |
|---|---|---|---|---|---|---|---|---|---|---|---|
| CS | -1.95 | -2.12 | -1.07 | -1.74 | -0.49 | -0.31 | -0.82 | 0.29 | -1.11 | 0.29 | 2.63 |
| Physics | -1.47 | -0.43 | -0.49 | -1.14 | 0.07 | -0.87 | 0.15 | -0.25 | -0.54 | -0.47 | 2.08 |
| Math | -0.80 | -1.18 | -1.43 | -0.68 | -0.58 | -1.13 | 0.59 | -1.63 | -1.37 | 1.16 | 2.23 |
| Stat | -3.23 | -1.60 | -1.04 | -2.69 | -0.44 | 2.17 | 0.24 | -0.91 | 0.39 | 0.80 | 3.93 |
| EESS | 0.00 | 5.88 | 2.50 | -3.65 | -0.58 | 0.84 | -1.34 | 1.03 | -0.80 | -0.06 | 2.49 |
| q-bio | -0.48 | -0.69 | 0.77 | -1.87 | 0.22 | -1.12 | 0.06 | -0.43 | -3.33 | 0.34 | 7.20 |
| q-fin | 6.53 | -4.14 | -5.02 | -0.29 | 4.92 | -2.64 | -0.26 | 2.60 | -7.16 | 6.13 | 6.35 |
| Econ | 0.00 | 0.00 | 25.66 | 2.05 | -1.54 | -4.15 | 1.41 | 0.34 | -2.20 | 7.78 | 1.03 |

**Table 5.** Year-over-year change in the within-category single-author share by category, first quarter (January to March), 2016 to 2026, in percentage points.

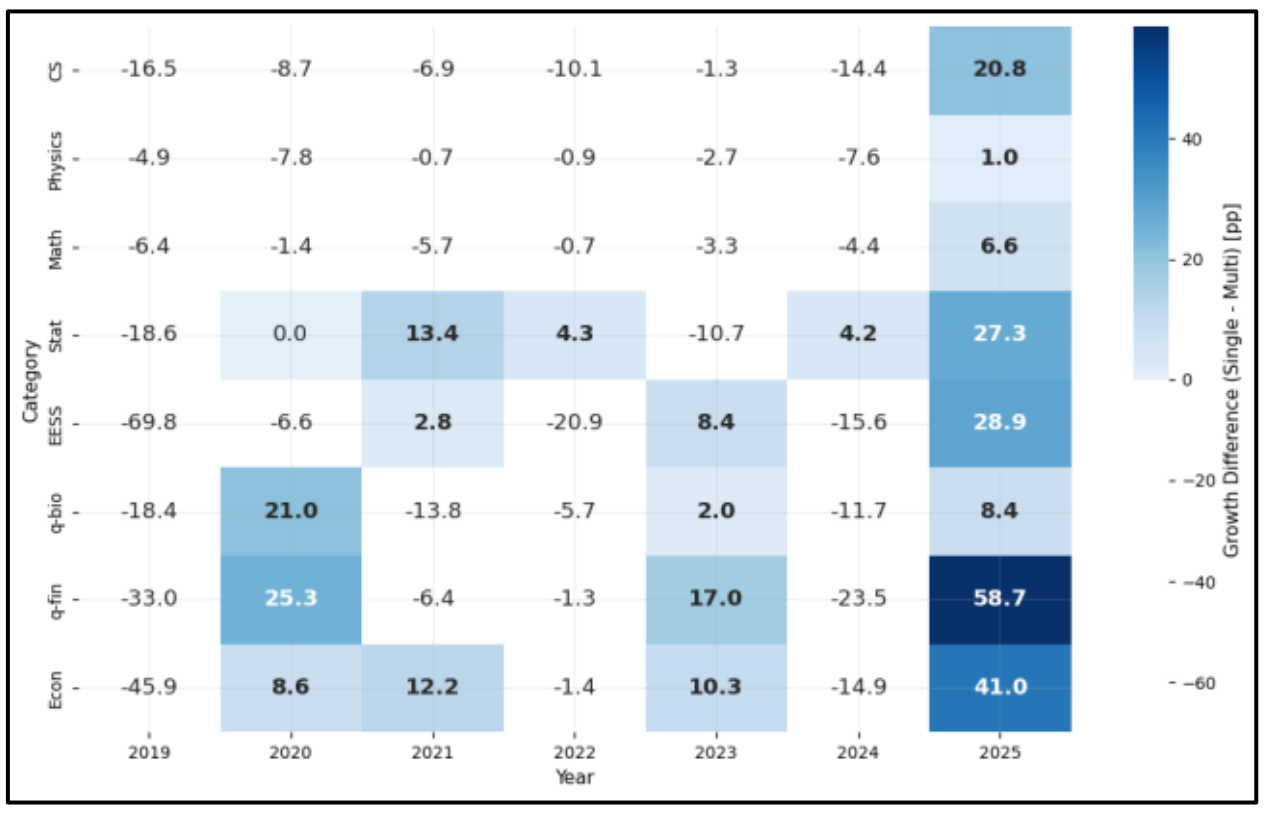

**Figure 6.** Difference in year-over-year growth between single- and multi-authored papers by category, annual, 2019 to 2025, in percentage points (bolded values indicate the years in which single-author growth exceeded multi-author growth).

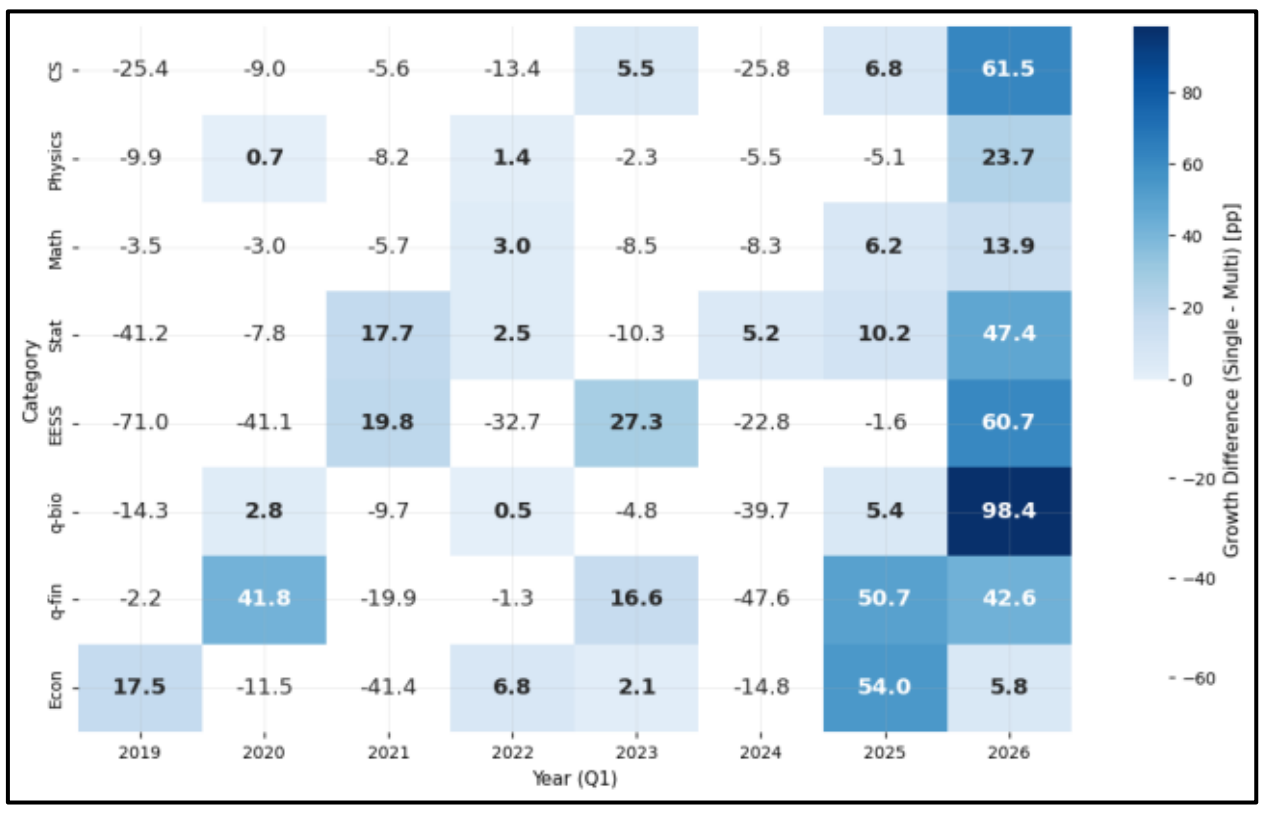

**Figure 7.** Difference in year-over-year growth between single- and multi-authored papers by category, first quarter (January to March), 2019 to 2026, in percentage points.

The third research question (RQ3) concerns whether the observed trends, and any recent departure from them, differ across the eight top-level arXiv categories. The within-category single-author share is computed for each category for each year between 2015 and 2025 and for the first quarter of each year between 2015 and 2026, and year-over-year changes are reported by category. Paralleling RQ1, year-over-year growth in single-authored papers is compared with that of multi-authored papers within each category, making it possible to assess whether the recent change, in which single-authored papers grew more quickly, is observable at the category level.

Tables 4 and 5 report the year-over-year change in the within-category single-author share, at the annual and first-quarter levels respectively. Among the categories with substantial submission volumes throughout the period, the within-category share fell in nearly every year from 2015 through 2024 at both levels. Between 2024 and 2025 at the annual level, and between the first quarter of 2025 and the first quarter of 2026, the within-category single-author share rose in every one of the eight top-level categories. The increases were observed across categories that differ widely in size and baseline single-author rates, indicating that the reversal is a general feature of the period rather than an artifact of any single field.

The growth comparison from RQ1, in which single-authored papers grew faster than multi-authored papers for the first time in 2025 at the aggregate level, holds within every category individually. Figures 6 and 7 present the difference in year-over-year growth between single- and multi-authored papers for each category and year, with bolded values marking the cells in which single-author growth exceeded multi-author growth. Several smaller categories had limited data before 2019: EESS and Econ did not yet exist as established categories on arXiv, while q-fin and q-bio had small submission volumes. The figures therefore display results from 2019 onward.

In 2025 at the annual level, and in 2026 at the first-quarter level, single-author growth exceeded multi-author growth in all eight categories. At the annual level, this occurred in at most four categories in any year from 2016 to 2024 (in 2023) and never in CS, Physics, or Math, the three largest categories. At the first-quarter level, it occurred in at most five categories in any year from 2016 to 2024 (in 2022) and in six in 2025. The 2025 annual and 2026 first-quarter results are the first since 2016 in which all eight categories simultaneously showed single-author growth above multi-author growth. These results therefore provide a per-category corroboration of the same pattern documented at the aggregate level in RQ1.

## 8. Validation of Author-List Stability Over Time

| Earlier snapshot | Later snapshot | Years elapsed | Solo → multi (count) | Multi → solo (count) | Solo → multi (%) | Multi → solo (%) | Solo papers still solo (%) | Multi papers still multi (%) | Solo → multi (% per year) | Multi → solo (% per year) |
|---|---|---|---|---|---|---|---|---|---|---|
| v72 | v124 | 1 | 256 | 83 | 0.058 | 0.005 | 99.94 | 99.99 | 0.058 | 0.005 |
| v72 | v175 | 2 | 375 | 122 | 0.086 | 0.008 | 99.91 | 99.99 | 0.043 | 0.004 |
| v72 | v227 | 3 | 448 | 159 | 0.103 | 0.010 | 99.90 | 99.99 | 0.034 | 0.003 |
| v72 | v281 | 4 | 511 | 176 | 0.117 | 0.011 | 99.88 | 99.99 | 0.029 | 0.003 |
| v124 | v175 | 1 | 305 | 79 | 0.066 | 0.004 | 99.93 | 99.99 | 0.066 | 0.004 |
| v124 | v227 | 2 | 437 | 141 | 0.095 | 0.008 | 99.91 | 99.99 | 0.048 | 0.004 |
| v124 | v281 | 3 | 524 | 166 | 0.114 | 0.009 | 99.89 | 99.99 | 0.038 | 0.003 |
| v175 | v227 | 1 | 286 | 112 | 0.059 | 0.006 | 99.94 | 99.99 | 0.059 | 0.006 |
| v175 | v281 | 2 | 427 | 159 | 0.088 | 0.008 | 99.91 | 99.99 | 0.044 | 0.004 |
| v227 | v281 | 1 | 349 | 126 | 0.068 | 0.006 | 99.93 | 99.99 | 0.068 | 0.006 |

**Table 6.** Rates of change in author counts and retention of single- and multi-author status between pairs of arXiv snapshots. Years elapsed are rounded to whole numbers.

Because the most recently posted papers in the dataset (those submitted in 2025 and early 2026) have had less time for their author lists to change than older papers, it is necessary to confirm that the recent increase in single authorship is not simply a result of this difference in elapsed time before any conclusion is drawn from the most recent years. The concern is that some papers initially posted with a single author may later have additional authors added and so become multi-authored, and that for papers posted close to the date of the most recent snapshot this process may not yet have run its course.

To measure how often author lists actually change after posting, the five snapshots described in Section 2 are compared directly. For each paper present in the most recent snapshot, its author count is traced back through every earlier snapshot in which it appears, making it possible to follow the same paper across up to four years. For each pair of snapshots, the number of papers that changed from single-authored to multi-authored, and the number that changed from multi-authored to single-authored, are computed and expressed both as totals and as rates per year of elapsed time. The measured rates are then used to assess how sensitive the recent findings are to the difference in elapsed time across papers.

The first finding is that author lists are highly stable after a paper is first posted. Table 6 reports, for every pair of snapshots, the rate at which single-authored papers gained one or more co-authors, the rate at which multi-authored papers were reduced to a single author, and the percentages of papers retaining their original status. Across all ten pairs, the rate at which single-authored papers gained co-authors was below 0.07 percentage points per year of elapsed time, and the rate at which multi-authored papers were reduced to a single author was below 0.01 percentage points per year. Of papers single-authored in the April 2022 snapshot, 99.88% were still single-authored in the April 2026 snapshot, four years later (only 511 out of 435,268 single-authored papers gained an author). The corresponding figure for multi-authored papers was 99.99%. Changes to author lists after the initial posting are therefore rare, and rare in a way that is roughly constant across the period rather than concentrated in any particular years.

Because these rates are so small, the bias they could introduce into the recent findings is negligible. Even if the first-quarter 2026 single-author share were adjusted downward by the amount of attrition expected over an additional year of elapsed time, using the highest measured rate of co-author addition from any pair of snapshots, the increase between the first quarter of 2025 and the first quarter of 2026 would remain at 2.19 percentage points, compared with the 2.20 observed. The recent rise in single authorship therefore cannot be explained by the limited time that recently posted papers have had for their author lists to change.

## 9. Validation of Results using OpenAlex

| Year | Total | Single share (%) | Single growth (%) | Multi growth (%) | Total Q1 | Single share Q1 (%) | Single growth Q1 (%) | Multi growth Q1 (%) |
|---|---|---|---|---|---|---|---|---|
| 2015 | 7,299,686 | 39.27 | - | - | 3,640,349 | 47.14 | - | - |
| 2016 | 7,369,968 | 35.08 | -9.81 | 7.93 | 3,647,791 | 39.84 | -15.30 | 14.03 |
| 2017 | 6,984,103 | 33.69 | -8.99 | -3.20 | 3,204,317 | 38.79 | -14.47 | -10.63 |
| 2018 | 6,894,456 | 32.58 | -4.54 | 0.37 | 2,919,200 | 38.81 | -8.86 | -8.92 |
| 2019 | 7,171,218 | 31.79 | 1.50 | 5.23 | 2,930,882 | 37.48 | -3.06 | 2.59 |
| 2020 | 7,301,671 | 29.98 | -4.00 | 4.53 | 2,908,002 | 36.21 | -4.14 | 1.24 |
| 2021 | 6,728,597 | 25.48 | -21.67 | -1.93 | 2,509,131 | 30.65 | -26.96 | -6.20 |
| 2022 | 6,138,569 | 22.52 | -19.36 | -5.15 | 2,034,275 | 24.85 | -34.25 | -12.15 |
| 2023 | 6,415,299 | 22.44 | 4.14 | 4.62 | 2,125,030 | 25.99 | 9.24 | 2.88 |
| 2024 | 6,765,443 | 20.90 | -1.80 | 7.56 | 2,135,933 | 24.84 | -3.92 | 2.07 |
| 2025 | 7,979,834 | 22.95 | 29.52 | 14.89 | 2,381,705 | 26.72 | 19.94 | 8.72 |
| 2026* | 3,880,584 | 30.28 | - | - | 2,548,720 | 30.55 | 22.34 | 1.42 |

**Table 7.** Annual and first quarter single- and multi-authored OpenAlex submissions of article type, 2015–2026. *Partial data for 2026

One concern about the findings is that they rely solely on data from arXiv, which covers a specific set of categories that does not represent all of scientific research. To assess whether the findings can be observed in a different data source, the OpenAlex database was used to repeat the core analysis. The official OpenAlex API was used to retrieve all records of type “article” published between 2015 and May 2026 (the retrieval date), which excludes preprints and captures predominantly peer-reviewed publications. The same single-author and growth analyses were then computed on this data. Table 7 reports the results.

The findings are consistent with those obtained from arXiv: for the first time in the studied window at the annual level, the growth of single-authored articles exceeded that of multi-authored articles in 2025. A similar pattern is observed in the first-quarter comparison between 2025 and 2026, in which single-authored articles grew by 22.34% and multi-authored articles by 1.42%, although the first-quarter growth of single-authored articles was also higher for three earlier years (2018, 2023, and 2025). The single-author share itself rose from 26.72% in Q1 2025 to 30.55% in Q1 2026. While the OpenAlex and arXiv data should not be expected to agree exactly given their different coverage, metadata, and known quality limitations in OpenAlex, the qualitative consistency of the two analyses reduces the likelihood that the reversal documented here is specific to arXiv's platform or metadata practices.

It is important to note that although OpenAlex offers far broader coverage, concerns have been raised about the quality and consistency of its data (Alonso-Álvarez & Van Eck, 2025; Forchino & Torres-Salinas, 2026). OpenAlex may also suffer from issues such as indexing lag, incomplete coverage of recent publications, and duplicate records arising from overlapping sources. These concerns are particularly relevant here: the study focuses on possible recent shifts in single authorship, and the most recent records, which are central to the analysis, are also the most likely to contain errors that have not yet been identified and corrected. For this reason, OpenAlex serves only as a robustness check on the main arXiv-based findings, not as a primary data source. Future research could revisit OpenAlex once errors in its most recent records have had time to be identified and corrected. A further step would be a like-for-like comparison: restricting OpenAlex to the fields covered by the arXiv categories studied here and testing whether the two sources show the same patterns.

## 10. Validation of Results using October 2026 Release

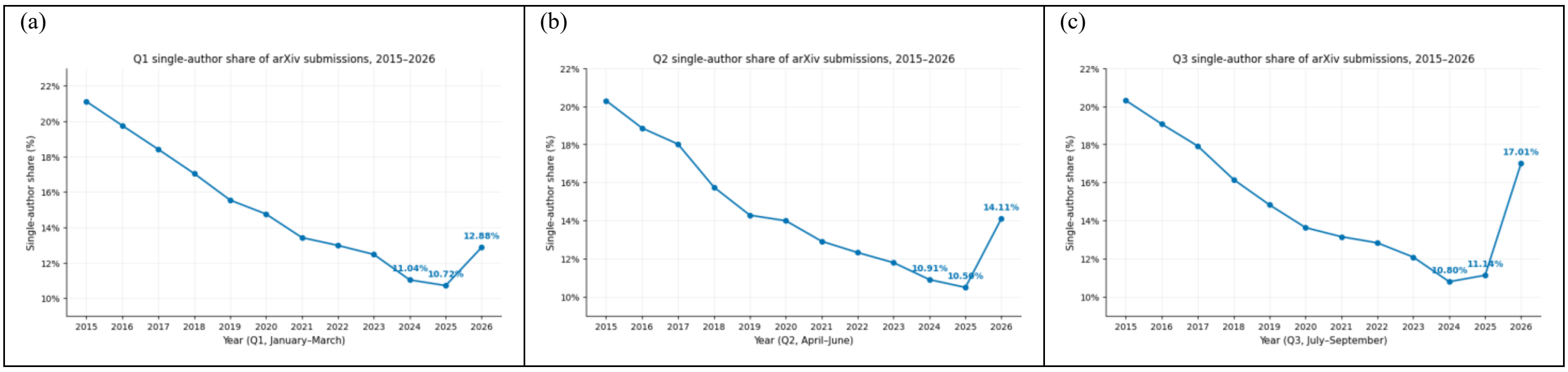

**Figure 8.** Single-author share of arXiv submissions in the October 2026 release (version 306), 2015–2026: (a) Q1 (January–March), (b) Q2 (April–June), (c) Q3 (July–September).

The initial analysis concluded with version 281, released in April 2026. To validate those results, the work was repeated using version 306, released in October 2026. The objective is twofold: to confirm the robustness of the earlier findings, and to assess how the observed patterns evolve now that nine months of 2026 data are available. The replication proceeds in two parts. First, the entire preceding analysis is reproduced on version 306 data. This step addresses potential concerns such as shifts in author lists over time, for instance, originally single-authored papers that subsequently acquired additional authors. Second, because the October 2026 release covers the first three quarters of 2026, Q2 and Q3 are treated as additional units of analysis, and the change in the single-authored share is quantified first for Q2 and then for Q3.

Repeating the earlier analysis in full confirmed the original findings. The share of single-authored papers in 2026, once the additional months are included, in fact rose from the previously reported 12.99% (January to early April) to 14.83% (January to September). For the Q1 data, the share declined only marginally, from 12.93% to 12.88%. These results corroborate the earlier findings and indicate that the observed trend is persistent.

The final analysis extended the Q1 focus of the earlier work to Q2 and Q3. The results, presented in Figure 8, indicate that the identified trend strengthens further in these two quarters. The share of single-authored papers in Q2 2026 rose to 14.11%, up from 10.50% in Q2 2025, and exceeded the Q1 2026 level of 12.88%. The increase in Q3 was more pronounced: 17.01% of papers were single-authored, compared with 11.14% in Q3 2025. In summary, these findings corroborate the earlier results: the gap relative to the same quarter of 2025 widened from 3.61 percentage points in Q2 to 5.87 in Q3.

It is important to note, however, that on October 1, 2026, arXiv updated its submission rate-limit policy (Boboris, 2026). Each submitter may now post at most two new submissions per calendar month. Future analyses may therefore observe a decline in single-authored submissions attributable to this change. Because a submission counts only against the quota of the person who submits it, not against those of the other co-authors, the policy may also encourage collaboration: a team can spread its papers across several submitters, whereas a single author cannot.

## 11. Conclusion

In this paper, a dataset of papers posted to arXiv between 2015 and early 2026 was analyzed to identify changes in the share of single-authored papers over time, with a focus on determining whether the long-established decline in single authorship has continued or shifted in the most recent period. The results showed a steady decline in the single-author share from 2015 through 2024, followed by a reversal in 2025 and early 2026 in which the single-author share rose at both the annual and first-quarter levels and in which the number of single-authored papers grew more quickly than the number of multi-authored papers for the first time in the studied window. Further analysis showed that the recent reversal was observed across all eight top-level arXiv categories, reflected genuine changes within fields rather than shifts in the composition of arXiv submissions across fields, and was not an artifact of incomplete or evolving author metadata in recently posted papers.

This study was motivated by the rapid spread of AI tools and by the hypothesis that they allow more researchers to work independently. An important limitation, however, is that the analysis documents the recent reversal in single authorship without testing whether AI caused it. Future research could test this link directly: surveys could ask researchers whether AI tools influence their decision to work alone, and full-text analysis could compare estimated AI use in single-authored and multi-authored papers. Another direction would be to track individual researchers over time and see whether those who adopt AI tools become more likely to publish alone. In the end, this study contributes to the ongoing body of literature on the evolution of authorship and team size in scientific publishing, and establishes an empirical pattern that warrants further investigation.

## Declaration of generative AI and AI-assisted technologies in the manuscript preparation process.

During the preparation of this work, the author used Opus 4.7 and Opus 5.5 from Anthropic to aid in writing, general preparation, and code writing for the study. The author also used Gemini 3.1 Pro from Google to suggest visualization and aid in programming. After using these tools/services, the author reviewed and edited the content as needed and takes full responsibility for the content of the published article.

## Appendix:

### A. Team/Solo Author Classifier

| arXiv_id | Authors | Manual Annotation | Classifier |
|---|---|---|---|
| hep-th/0508179 | Jorgen Rasmussen | Solo | Solo |
| 2310.10057 | Toshihiro Nakanishi | Solo | Solo |
| nucl-ex/0406027 | Burt Holzman (for the PHOBOS Collaboration) | Team | Team |
| 1304.6386 | ATLAS collaboration | Team | Team |
| 2009.12628 | CMS Collaboration | Team | Team |
| 2404.07708 | Valery I. Kovalchuk | Solo | Solo |
| 2311.08231 | The HIKE Collaboration | Team | Team |
| cond-mat/0607020 | Joachim Ankerhold | Solo | Solo |
| quant-ph/0612217 | Alexander Jurisch | Solo | Solo |
| 1311.2332 | Benjamin Antieau | Solo | Solo |

**Table A.1.** Examples from the validation sample: manual annotations and classifier labels

| Year | Papers | Single-authored (original) | Single-authored (original) share (%) | Single-authored (strict) | Single-authored (strict) share (%) | Reclassified | Reclassified share (% of single-authored) |
|---|---|---|---|---|---|---|---|
| 2015 | 105,130 | 21,637 | 20.58 | 21,054 | 20.03 | 583 | 2.69 |
| 2016 | 113,440 | 21,830 | 19.24 | 21,346 | 18.82 | 484 | 2.22 |
| 2017 | 123,781 | 22,293 | 18.01 | 21,688 | 17.52 | 605 | 2.71 |
| 2018 | 140,377 | 22,811 | 16.25 | 22,265 | 15.86 | 546 | 2.39 |
| 2019 | 155,917 | 23,180 | 14.87 | 22,658 | 14.53 | 522 | 2.25 |
| 2020 | 178,275 | 24,763 | 13.89 | 24,408 | 13.69 | 355 | 1.43 |
| 2021 | 181,599 | 23,895 | 13.16 | 23,484 | 12.93 | 411 | 1.72 |
| 2022 | 185,987 | 23,566 | 12.67 | 23,117 | 12.43 | 449 | 1.91 |
| 2023 | 209,223 | 25,149 | 12.02 | 24,681 | 11.80 | 468 | 1.86 |
| 2024 | 243,690 | 26,581 | 10.91 | 26,108 | 10.71 | 473 | 1.78 |
| 2025 | 284,161 | 31,850 | 11.21 | 31,368 | 11.04 | 482 | 1.51 |
| 2026* | 85,475 | 11,101 | 12.99 | 10,958 | 12.82 | 143 | 1.29 |

**Table A.2.** Single-authored arXiv papers under the original and strict definitions, 2015–2026. *Partial data for 2026

This appendix supplements the description of the classifier in Section 3. The classifier is applied to every paper whose parsed author list contains a single entry, and labels the paper as a team paper if either of two rules is satisfied: (1) the author entry contains the word “collaboration” or “consortium”; or (2) the entry contains “for” or "on behalf of," followed later by "collaboration," "group," "team," or "consortium." The rules are implemented in Python as the following two regular expressions, both of which ignore capitalization:

```
import re
COLLAB_NAME_RE = re.compile(r'\b(collaboration|consortium)\b', re.IGNORECASE)
REPR_RE = re.compile(r'(for the|on behalf of|for)\b.*\b(collaboration|group|team|consortium)\b', re.IGNORECASE)
```

Regarding the performance of the classifier, because the validation sample over-represented papers the classifier labeled as collaborations, it contains 100 team papers and 150 truly single-authored papers. This split does not reflect their prevalence in the full dataset. Treating team papers as the positive class, the classifier produced 100 true positives, 150 true negatives, and no false positives or false negatives, so precision, recall, and accuracy are all 100% (95% confidence interval for accuracy: 98.5–100%). Perfect agreement in a sample of 250 does not rule out rare errors in the full dataset, but given the small effect of reclassification (Table A.2), such errors are unlikely to affect the findings.

Table A.1 shows ten papers from the validation sample with their manual annotations and classifier labels. Table A.2 reports the annual counts and shares of single-authored papers under the original and strict definitions. The objective of this table is to show the effect of reclassification. Under both definitions, the share reaches its lowest point in 2024 and then rises: under the strict definition, from 10.71% to 11.04% in 2025 and 12.82% in the available months of 2026, compared with 10.91%, 11.21%, and 12.99% under the original definition.

## B. Statistical Tests of Year-over-Year Changes

| Comparison | Change (pp) | Pooled SE (pp) | z | p | 95% CI (pp) | Significant (α = 0.05 / Bonferroni) |
|---|---|---|---|---|---|---|
| 2015 → 2016 | −1.34 | 0.17 | −7.83 | < 0.0001 | −1.67 to −1.00 | Yes / Yes |
| 2016 → 2017 | −1.23 | 0.16 | −7.71 | < 0.0001 | −1.55 to −0.92 | Yes / Yes |
| 2017 → 2018 | −1.76 | 0.15 | −12.00 | < 0.0001 | −2.05 to −1.47 | Yes / Yes |
| 2018 → 2019 | −1.38 | 0.13 | −10.38 | < 0.0001 | −1.64 to −1.12 | Yes / Yes |
| 2019 → 2020 | −0.98 | 0.12 | −8.03 | < 0.0001 | −1.22 to −0.74 | Yes / Yes |
| 2020 → 2021 | −0.73 | 0.11 | −6.42 | < 0.0001 | −0.96 to −0.51 | Yes / Yes |
| 2021 → 2022 | −0.49 | 0.11 | −4.41 | < 0.0001 | −0.70 to −0.27 | Yes / Yes |
| 2022 → 2023 | −0.65 | 0.10 | −6.21 | < 0.0001 | −0.86 to −0.44 | Yes / Yes |
| 2023 → 2024 | −1.11 | 0.09 | −11.73 | < 0.0001 | −1.30 to −0.93 | Yes / Yes |
| 2024 → 2025 | +0.30 | 0.09 | +3.47 | 0.0005 | 0.13 to 0.47 | Yes / Yes |

**Table B.1.** Two-proportion z-tests of year-over-year changes in the annual single-author share, 2015–2025.

| Comparison | Change (pp) | Pooled SE (pp) | z | p | 95% CI (pp) | Significant (α = 0.05 / Bonferroni) |
|---|---|---|---|---|---|---|
| 2015 → 2016 | −1.36 | 0.35 | −3.84 | 0.0001 | −2.05 to −0.66 | Yes / Yes |
| 2016 → 2017 | −1.34 | 0.33 | −4.06 | < 0.0001 | −2.00 to −0.69 | Yes / Yes |
| 2017 → 2018 | −1.38 | 0.31 | −4.48 | < 0.0001 | −1.98 to −0.77 | Yes / Yes |
| 2018 → 2019 | −1.49 | 0.28 | −5.30 | < 0.0001 | −2.05 to −0.94 | Yes / Yes |
| 2019 → 2020 | −0.79 | 0.26 | −3.02 | 0.0026 | −1.30 to −0.28 | Yes / Yes |
| 2020 → 2021 | −1.33 | 0.24 | −5.51 | < 0.0001 | −1.80 to −0.86 | Yes / Yes |
| 2021 → 2022 | −0.44 | 0.23 | −1.95 | 0.052 | −0.89 to 0.00 | No / No |
| 2022 → 2023 | −0.51 | 0.22 | −2.32 | 0.021 | −0.94 to −0.08 | Yes / No |
| 2023 → 2024 | −1.43 | 0.20 | −7.17 | < 0.0001 | −1.82 to −1.04 | Yes / Yes |
| 2024 → 2025 | −0.32 | 0.18 | −1.80 | 0.072 | −0.67 to 0.03 | No / No |
| 2025 → 2026 | +2.20 | 0.17 | +12.78 | < 0.0001 | 1.87 to 2.54 | Yes / Yes |

**Table B.2.** Two-proportion z-tests of year-over-year changes in the first-quarter (January–March) single-author share, 2015–2026.

This appendix supplements the results in Section 5. The single-author share in each year was compared with that of the previous year using a two-proportion z-test on the counts in Table 2 (annual) and Table 3 (first quarter). The test uses a pooled standard error, which estimates a single common share from both years. This is appropriate because the test evaluates the null hypothesis that there is no change, under which both years have the same underlying share. The 95% confidence intervals use unpooled standard errors, since they estimate the size of the change without assuming that it is zero. Significance is also assessed against Bonferroni-corrected thresholds of 0.005 for the ten annual comparisons and 0.0045 for the eleven first-quarter comparisons.

At the annual level (Table B.1), every change is statistically significant, including the 2025 increase. The partial-year 2026 figure is not tested, as it is not comparable with full-year 2025. At the first-quarter level (Table B.2), the 2026 increase is highly significant, while the smaller declines in 2022 and 2025 are not, and the 2023 decline is not significant after correction. The first-quarter share was therefore effectively flat between 2024 and 2025 before rising sharply in 2026.